\documentclass{article}%
\usepackage{amsmath}%
\usepackage{amsfonts}%
\usepackage{amssymb}%
\usepackage{graphicx}
\begin{document}

\author{Sawa Manoff\\Bulgarian Academy of Sciences\\Institute for Nuclear Research and Nuclear Energy\\Department of Theoretical Physics\\Laboratory of Solitons, Coherency, and Geometry\\1784 Sofia - Bulgaria}
\date{E-mail address: smanov@inrne.bas.bg}
\title{Doppler effect and Hubble effect in different models of space-time in the case
of auto-parallel motion of the observer}
\maketitle

\begin{abstract}
Doppler effect and Hubble effect in different models of space-time in the case
of auto-parallel motion of the observer are considered. The Doppler effect and
shift frequency parameter are specialized for the case of auto-parallel motion
of the observer. The Hubble effect and shift frequency parameter are
considered for the same case. It is shown that by the use of the variation of
the shift frequency parameter during a time period, considered locally in the
proper frame of reference of an observer, one can directly determine the
centrifugal (centripetal) relative velocity \ and acceleration as well as the
Coriolis relative velocity and acceleration of an astronomical object moving
relatively to the observer. All results are obtained on purely kinematic basis
without taking into account the dynamic reasons for the considered effect. 


\end{abstract}
\tableofcontents

\section{\bigskip Introduction}

1. Modern problems of relativistic astrophysics as well as of relativistic
physics (dark matter, dark energy, evolution of the universe, measurement of
velocities of moving objects etc.) are related to the propagation of signals
in space or in space-time \cite{Weinberg}, \cite{Unzicker}. The basis of
experimental data received as results of observations of the Doppler effect or
of the Hubble effect gives rise to theoretical considerations about the
theoretical status of effects related to detecting signals from emitters
moving relatively to observers carrying detectors in their laboratories.

2. In the classical (non-quantum) field theories different models of
space-time have been used for describing the physical phenomena and their
evolution. The $3$-dimensional Euclidean space $E_{3}$ is the physical space
used as the space basis of classical mechanics \cite{Javorski}. The
$4$-dimensional (flat) Minkowskian space $\overline{M}_{4}$ is used as the
model of space-time in special relativity \cite{Tonnelat}. The (pseudo)
Riemannian spaces $V_{4}$ without torsion are considered as models of
space-time in general relativity \cite{Anderson}, \cite{Misner}. In
theoretical gravitational physics (pseudo) Riemannian spaces without torsion
as well as (pseudo) Riemannian spaces $U_{4}$ with torsion are proposed as
space-time grounds for new gravitational theories. To the most sophisticated
models of space-time belong the spaces with one affine connection and metrics
[$(L_{n},g)$-spaces] and the spaces with affine connections and metrics
[$(\overline{L}_{n},g)$-spaces].

The spaces with one affine connection and metrics [$(L_{n},g)$-spaces] have
affine connections whose components differ only by sign for contravariant and
covariant tensor fields over a differentiable manifold $M$ with $\dim~M=n$.

The spaces with affine connections and metrics have affine connections whose
components differ not only by sign for contravariant and covariant tensor
fields over a differentiable manifold $M$ with $\dim~M=n$.

3. Recently, it has been shown \cite{Manoff-1}, \cite{Manoff-2} that every
differentiable manifold $M$ ($dimM=n$) with affine connections and metrics
[$(\overline{L}_{n},g)$-spaces] \cite{Manoff-3} could be used as a model of
space-time for the following reasons:

\begin{itemize}
\item The equivalence principle (related to the vanishing of the components of
an affine connection at a point or on a curve) holds in $(\overline{L}_{n}%
,g)$-spaces \cite{Iliev-1}$\div$\cite{Iliev-1b}.

\item $(\overline{L}_{n},g)$-spaces have structures similar to these in
(pseudo) Riemannian spaces without torsion [$V_{n}$-spaces] allowing for
description of dynamic systems and the gravitational interaction
\cite{Manoff-2}.

\item Fermi-Walker transports and conformal transports exist in $(\overline
{L}_{n},g)$-spaces as generalizations of these types of transports in $V_{n}%
$-spaces \cite{Manoff-5}, \cite{Manoff-6}.

\item A Lorentz basis and a light cone could not be deformed in $(\overline
{L}_{n},g)$-spaces as it is the case in $V_{n}$-spaces.

\item All kinematic characteristics related to the notions of relative
velocity and of relative acceleration could be worked out in $(\overline
{L}_{n},g)$-spaces without changing their physical interpretations in $V_{n}%
$-spaces \cite{Manoff-2}, \cite{Manoff-7}$\div$\cite{Manoff-8a}.

\item $(\overline{L}_{n},g)$-spaces include all types of spaces with affine
connections and metrics used until now as models of space-time.
\end{itemize}

4. If a $(\overline{L}_{n},g)$-space \ could be used as a model of space or of
space-time the question arises how signals could propagate in a space-time
described by a $(\overline{L}_{n},g)$-space. The answer of this question has
been given in \cite{Manoff-8a}, \cite{Manoff-8}, and \cite{Manoff-9}. By that
the signals are described by means of null (isotropic) vector fields. A signal
could be defined as a periodical process transferred by an emitter and
received by an observer (detector) \cite{Manoff-8}. A wave front could be
considered as a signal characterized by its wave vector (null vector) as it is
the case in the geometrical optics in a $V_{n}$-space \cite{Stephani}. All
results are obtained on \textit{purely kinematic basis} (s. \cite{Manoff-8a},
\cite{Manoff-12}, \cite{Manoff-11}) without taking into account the dynamic
reasons for the considered effect.

On the basis of the general results in the previous papers we can draw a rough
scheme of the relations between the kinematic characteristics of the relative
velocity and relative acceleration on the one side, and the Doppler effect and
the Hubble effect on the other. Here the following abbreviations are used:

CM - classical mechanics

SRT - special relativity theory

GRT - general relativity theory

CRT - classical relativity theory \cite{Manoff-1}, \cite{Manoff-2}.

%

%TCIMACRO{\FRAME{ftbpF}{10.0034cm}{14.0298cm}{0pt}{}{}{sygnals.jpg}%
%{\special{ language "Scientific Word";  type "GRAPHIC";  display "USEDEF";
%valid_file "F";  width 10.0034cm;  height 14.0298cm;  depth 0pt;
%original-width 19.2358cm;  original-height 25.1975cm;  cropleft "0";
%croptop "1";  cropright "1";  cropbottom "0";
%filename 'C:/swp40/Docs/Sygnals.jpg';file-properties "XNPEU";}}}%
%BeginExpansion
\begin{figure}
[ptb]
\begin{center}
\includegraphics[
natheight=25.197500cm,
natwidth=19.235800cm,
height=14.0298cm,
width=10.0034cm
]%
{C: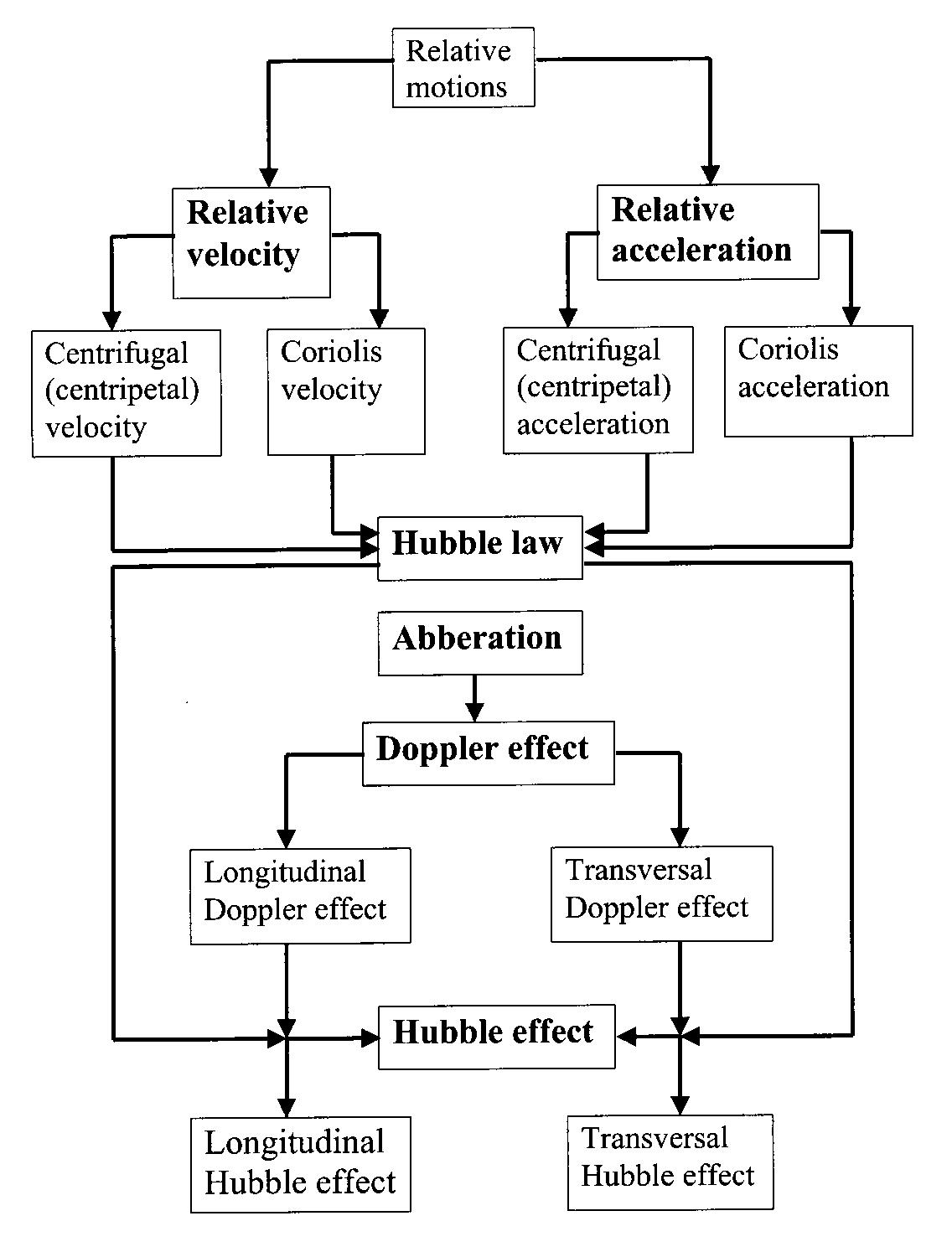}%
\end{center}
\end{figure}
%EndExpansion

\bigskip

5. The considerations of the Doppler effect and of the Hubble effect show that
the Doppler effect is derived in the physical theories (with exception of
general relativity) as a result of the relative motion of an observer and an
emitter, sending signals to the observer, from point of view of the proper
frame of reference of the observer and its relations to the proper frame of
reference of the emitter. On the other side, the Hubble effect could be
considered as a result of the Doppler effect and the Hubble law assumed to be
valid in the corresponding physical theory. In a rough scheme the relations
between Doppler effect, \ Hubble effect, and Hubble law could be represented
as follows:

\medskip%

\begin{tabular}
[c]{llllllll}
& {\small Relative motion} &  & {\small Doppler effect} &  & {\small Hubble
effect} &  & {\small Hubble law}\\
& {\small characterized} &  & {\small characterized}\textit{\ } &  &
{\small characterized} &  & {\small characterized}\\
& {\tiny \ }{\small by} &  & {\small as} &  & {\small as} &  & \\
{\small CM} & {\small constant } & $\Rightarrow$ & corollary & $\rightarrow$ &
corollary & $\Leftarrow$ & by definition\\
& {\small relative velocity} &  &  &  &  &  & \\
{\small SRT} & {\small constant } & $\Rightarrow$ & corollary & $\rightarrow$
& corollary & $\Leftarrow$ & by definition\\
& {\small relative velocity} &  &  &  &  &  & \\
{\small GRT} & {\small change of the} & $\Rightarrow$ & corollary &
$\rightarrow$ & corollary & $\Leftarrow$ & by definition\\
& {\small metrics of space-time} &  &  &  &  &  & of the metrics\\
{\small CRT} & {\small relative velocity and} & $\Rightarrow$ & corollary &
$\rightarrow$ & corollary & $\leftarrow$ & as corollary\\
& {\small relative acceleration} & $\Downarrow$ &  &  &  &  & $\Uparrow$\\
&  & $\Rightarrow$ & $\longrightarrow$ & $\rightarrow$ & $\longrightarrow$ &
$\rightarrow$ & $\Uparrow$%
\end{tabular}

6. Since the Doppler effect and the Hubble effect as kinematic effects could
be described by different theoretical schemes and models of space-time the
rich mathematical tools of the spaces with affine connections and metrics,
considered as models of space-time, are used for description of both the
effects. The aim has been to work out a theoretical model of the Doppler
effect and of the Hubble effect as corollaries only of the relative motion
between emitter and observer determined by the kinematic characteristics of
the relative velocity and the relative acceleration \ between emitter and
observer from point of view of the proper frame of reference of the observer.
For this task the $(n-1)+1$ formalism has been used related to the world line
of an observer and its corresponding $n-1$ dimensional sub space interpreted
as the observed space in the proper frame of the observer \cite{Manoff-12}.

7. Our task in the present paper is to investigate the Doppler effect and the
Hubble effect in the case of an auto-parallel motion of the observer and its
influence on the frequency shift parameter. In section 2 the Doppler effect
and shift frequency parameter are considered in the case of an auto-parallel
motion of the observer. In Section 3 the Hubble effect and shift frequency
parameter are considered for the same case. It is shown that by the use of the
shift frequency parameter, considered locally in the proper frame of reference
of an observer \cite{Manoff-12}, we can directly determine the centrifugal
(centripetal) relative velocity \ and acceleration as well as the Coriolis
relative velocity and acceleration. Section 4 comprises some concluding
remarks. Most of the details and derivation omitted in this paper could be
found in \cite{Manoff-8} and in \cite{Manoff-9}.

\subsection{Abbreviation and symbols}

\begin{itemize}
\item The vector field $u$ is the velocity vector field of an observer: $u\in
T(M)$, $\dim~M=n$, $n=4$.

\item The contravariant vector field $v_{z}=\mp l_{v_{z}}\cdot n_{\perp
}=H\cdot l_{\xi_{\perp}}\cdot n_{\perp}=H\cdot\xi_{\perp}$ is orthogonal to
$u$ and collinear to $\xi_{\perp}$. It is called centrifugal (centripetal)
relative velocity.

\item The function $H=H(\tau)$ is called Hubble function.

\item The contravariant vector field $a_{z}=\mp l_{a_{z}}\cdot n_{\perp
}=\overline{q}\cdot l_{\xi_{\perp}}\cdot n_{\perp}=\overline{q}\cdot\xi
_{\perp}$ is orthogonal to $u$ and collinear to $\xi_{\perp}$. It is called
centrifugal (centripetal) relative acceleration.

\item The function $\overline{q}=\overline{q}(\tau)$ is called acceleration
function (parameter).

\item The contravariant vector field $v_{\eta c}=\mp l_{v_{\eta c}}\cdot
m_{\perp}=\overline{H}_{c}\cdot l_{\xi_{\perp}}\cdot m_{\perp}=\overline
{H}_{c}\cdot\eta_{\perp}$ is orthogonal to $u$ and to $\xi_{\perp}$. It is
called Coriolis relative velocity.

\item The function $\overline{H}_{c}=\overline{H}_{c}(\tau)$ is called
Coriolis Hubble function.{}

\item The contravariant vector field $a_{\eta_{c}}=\mp l_{a_{\eta c}}\cdot
m_{\perp}=\overline{q}_{\eta c}\cdot l_{\xi_{\perp}}\cdot m_{\perp}%
=\overline{q}_{\eta c}\cdot\eta_{\perp}$ is orthogonal to $u$ and to
$\xi_{\perp}$. It is called Coriolis relative acceleration. 

\item The fuction $\overline{q}_{\eta c}=\overline{q}_{\eta c}(\tau)$ is
called Coriolis acceleration function (parameter).

\item $\overline{\omega}$ is the frequency of a signal emitted by an emitter
at a time $\tau-d\tau$ of the proper time of the observer.

\item $\omega$ is the frequency of a signal detected by the observer at a time
$\tau$ of the porper time of the observer.

\item $d\tau$ is the time interval in the proper frame of reference of the
observer for which the signal propagates from the emitter to the observer
(detector) at a space distance $dl$ in the proper frame of reference of the observer.
\end{itemize}

\section{Doppler effect and shift frequency parameter in the case of an
auto-parallel motion of the observer}

It has been shown \cite{Manoff-8}, \cite{Manoff-9} that in a $(\overline
{L}_{n},g)$-space longitudinal (standard) and transversal Doppler effects
could appear when signals are propagating from an emitter to an observer
(detector) moving relatively to each other.

\subsubsection{Longitudinal (standard) Doppler effect and the shift frequency
parameter}

1. Let us now consider the shift frequency parameter when the observer's world
line is an auto-parallel trajectory, i.e. when the velocity vector $u$ of the
observer fulfills the equation%
\[
\nabla_{u}u=f\cdot.u\text{ \ \ \ \ \thinspace, \ \ \ \ \ \ }f\in C^{\infty
}(M)\text{ .}
\]

Then, because of $\overline{g}[h_{u}(u)]=0$,%
\[
a_{\perp}=\overline{g}[h_{u}(a)]=f\cdot\overline{g}[h_{u}(u)]=0\text{ \ \ \ ,}%
\]%
\begin{align*}
(\nabla_{u}a)_{\perp} &  =\overline{g}[h_{u}(\nabla_{u}(f\cdot u))]=\overline
{g}[h_{u}((uf)\cdot u+f\cdot a)]=\\
&  =(uf)\cdot\overline{g}[h_{u}(u)]+f\cdot\overline{g}[h_{u}(a)]=0\text{
\ \ \ \ \ .}%
\end{align*}

Let a signal with frequency $\overline{\omega}$ is emitted by an emitter
\cite{Manoff-8} at the time $\tau-d\tau$ and is received by an observer
(detector) at the time $\tau$.

Since $\overline{\omega}=\omega(\tau-d\tau)$ and $\omega=\omega(\tau)$, we can
expand $\overline{\omega}$ in a Teylor row up to the second order of $d\tau$%
\[
\overline{\omega}=\omega(\tau-d\tau)\approx\omega(\tau)-\frac{d\omega}{d\tau
}_{\mid\tau}\cdot d\tau+\frac{1}{2}\cdot\frac{d^{2}\omega}{d\tau^{2}}%
_{\mid\tau}\cdot d\tau^{2}+O(d\tau)\text{ \ \ .}
\]

Then%
\[
d\omega=\overline{\omega}-\omega\approx-\frac{d\omega}{d\tau}\cdot d\tau
+\frac{1}{2}\cdot\frac{d^{2}\omega}{d\tau^{2}}\cdot d\tau^{2}\text{ \ \ ,}
\]%
\[
\frac{d\omega}{\omega}=\frac{\overline{\omega}-\omega}{\omega}\approx-\frac
{1}{\omega}\cdot\frac{d\omega}{d\tau}\cdot d\tau+\frac{1}{2}\cdot\frac
{1}{\omega}\cdot\frac{d^{2}\omega}{d\tau^{2}}\cdot d\tau^{2}\text{ \ \ \ \ .}
\]

On the other side, the general results in the case of an auto-parallel motion
of the observer read \cite{Manoff-8}, \cite{Manoff-9}%
\begin{align*}
\frac{\overline{\omega}-\omega}{\omega} &  =\frac{d\omega}{\omega}=\omega
\cdot(1-\frac{\overline{l}_{v_{z}}}{l_{\xi_{\perp}}}+\frac{1}{2}\cdot\frac
{dl}{l_{u}^{2}}\cdot\overline{l}_{a_{z}})=\\
&  =\omega\cdot(1-\frac{dl}{l_{u}}\cdot\frac{l_{v_{z}}}{l_{\xi_{\perp}}}%
+\frac{1}{2}\cdot\frac{dl^{2}}{l_{u}^{2}}\cdot\frac{l_{a_{z}}}{l_{\xi_{\perp}%
}})\text{ \ \ \ \ .}%
\end{align*}

Since%
\[
\frac{dl}{l_{u}}=d\tau\text{ \ \ \ \ , \ \ \ \ \ \ \ \ }\frac{dl^{2}}%
{l_{u}^{2}}=d\tau^{2}\text{ \ \ \ ,}
\]
we obtain the relations%
\[
\overline{\omega}=\omega\cdot(1-\frac{l_{v_{z}}}{l_{\xi_{\perp}}}\cdot
d\tau+\frac{1}{2}\cdot\frac{l_{a_{z}}}{l_{\xi_{\perp}}}\cdot d\tau^{2})\text{
\ \ \ \ \ .}
\]

2. If we consider only infinitesimal changes of the frequency $\omega$ for the
time interval $d\tau$, i.e. if $d\omega=\overline{\omega}-\omega$, we can
express the shift parameter $z=(\overline{\omega}-\omega)/\omega$ as an
infinitesimal quantity%
\begin{align*}
z  &  =\frac{d\omega}{\omega}=d(\log~\omega)=\\
&  =d\overline{z}=-\frac{1}{\omega}\cdot\frac{d\omega}{d\tau}\cdot d\tau
+\frac{1}{2}\cdot\frac{1}{\omega}\cdot\frac{d^{2}\omega}{d\tau^{2}}\cdot
d\tau^{2}=\\
&  =-\frac{l_{v_{z}}}{l_{\xi_{\perp}}}\cdot d\tau+\frac{1}{2}\cdot
\frac{l_{a_{z}}}{l_{\xi_{\perp}}}\cdot d\tau^{2}\text{ \ \ .}%
\end{align*}

On the other side, $d\overline{z}$ could be considered as a differential of
the function $\overline{z}$ depending on the proper time $\tau$ of the
observer, i.e. $\overline{z}=\overline{z}(\tau)$. It is assumed that
$\overline{z}(\tau)$ has the necessary differentiability properties. The
function $\overline{z}$ at the point $\overline{z}(\tau-d\tau)$ could be
represented in Teylor row as
\[
\overline{z}(\tau-d\tau)=\overline{z}(\tau)-\frac{d\overline{z}}{d\tau}\cdot
d\tau+\frac{1}{2}\cdot\frac{d^{2}\overline{z}}{d\tau^{2}}\cdot d\tau
^{2}+O(d\tau)\text{ \ \ .\ \ }%
\]
Then $\overline{z}(\tau-d\tau)$ and $d\overline{z}=\overline{z}(\tau
-d\tau)-\overline{z}(\tau)$ could be written up to the second order of $d\tau$
respectively as%

\begin{align*}
\overline{z}(\tau-d\tau)  &  =\overline{z}(\tau)-\frac{d\overline{z}}{d\tau
}\cdot d\tau+\frac{1}{2}\cdot\frac{d^{2}\overline{z}}{d\tau^{2}}\cdot
d\tau^{2}\text{ \ \ \ ,}\\
d\overline{z}  &  =\overline{z}(\tau-d\tau)-\overline{z}(\tau)=\\
&  =-\frac{d\overline{z}}{d\tau}\cdot d\tau+\frac{1}{2}\cdot\frac
{d^{2}\overline{z}}{d\tau^{2}}\cdot d\tau^{2}=\\
&  =-\frac{1}{\omega}\cdot\frac{d\omega}{d\tau}\cdot d\tau+\frac{1}{2}%
\cdot\frac{1}{\omega}\cdot\frac{d^{2}\omega}{d\tau^{2}}\cdot d\tau^{2}=\\
&  =-\frac{l_{v_{z}}}{l_{\xi_{\perp}}}\cdot d\tau+\frac{1}{2}\cdot
\frac{l_{a_{z}}}{l_{\xi_{\perp}}}\cdot d\tau^{2}\text{ \ \ .}%
\end{align*}

The comparison of the coefficients before $d\tau$ and $d\tau^{2}$ in the last
(above) two expressions leads to the relations%
\[
\frac{d\overline{z}}{d\tau}=\frac{1}{\omega}\cdot\frac{d\omega}{d\tau}\text{
\ \ \ \ , \ \ \ \ \ \ }\frac{d^{2}\overline{z}}{d\tau^{2}}=\frac{1}{\omega
}\cdot\frac{d^{2}\omega}{d\tau^{2}}\text{ \ \ \ ,\ }
\]%
\[
\frac{d\overline{z}}{d\tau}=\frac{l_{v_{z}}}{l_{\xi_{\perp}}}\text{ \ \ \ \ ,
\ \ \ \ \ \ \ \ \ \ }\frac{d^{2}\overline{z}}{d\tau^{2}}=\frac{l_{a_{z}}%
}{l_{\xi_{\perp}}}\text{\ \ .}
\]

The vector $\xi_{\perp}$ could be chosen as a unit vector, i.e. $l_{\xi
_{\perp}}=1$, equal to the vector $n_{\perp}$ showing the direction to the
emitter from point of view of the observer. Then%
\[
\frac{d\overline{z}}{d\tau}=l_{v_{z}}\text{ \ \ \ \ \ \ ,
\ \ \ \ \ \ \ \ \ \ }\frac{d^{2}\overline{z}}{d\tau^{2}}=l_{a_{z}}\text{
\ \ \ \ \ .\ }%
\]

Therefore, if we can measure the change (variation) of the shift frequency
parameter $d\overline{z}$ in a time interval $d\tau$ we can find the
centrifugal (centripetal) relative velocity and centrifugal (centripetal)
relative acceleration of the emitter with respect to the observer. The above
relations appear as direct way for a check-up of the considered theoretical
scheme of the propagation of signals in spaces with affine connections and
metrics. On the other side, the explicit form of $l_{v_{z}}$ and $l_{a_{z}}$
as functions of the kinematic characteristics of the relative velocity and
relative acceleration could lead to conclusions of the properties of the
space-time model used for description of the physical phenomena. The same
relations could lead to more precise assessment of the Hubble function $H$ and
the acceleration function $\overline{q}$ at a given time.

\subsubsection{Transversal Doppler effect and the shift frequency parameter}

In analogous way we can find the relations between the absolute values of the
Coriolis relative velocity and the Coriolis relative acceleration and the
shift frequency parameter in the case of auto-parallel world line of the observer.

1. The relation between the frequency $\overline{\omega}$ of the emitted
signals and the frequency $\omega$ of the detected signals reads
\cite{Manoff-8}, \cite{Manoff-9}%
\begin{align*}
\overline{\omega} &  =\omega\cdot(1-\frac{dl}{l_{u}}\cdot\frac{l_{v_{\eta c}}%
}{l_{\xi_{\perp}}}+\frac{1}{2}\cdot\frac{dl^{2}}{l_{u}^{2}}\cdot
\frac{l_{a_{\eta c}}}{l_{\xi_{\perp}}})=\\
&  =\omega\cdot(1-\frac{l_{v_{\eta c}}}{l_{\xi_{\perp}}}\cdot d\tau+\frac
{1}{2}\cdot\frac{l_{a_{\eta c}}}{l_{\xi_{\perp}}}\cdot d\tau^{2})\text{
\ \ \ .}%
\end{align*}

The shift frequency parameter has the form%
\begin{align*}
z_{c}  &  =\frac{\overline{\omega}-\omega}{\omega}=-\frac{l_{v_{\eta c}}%
}{l_{\xi_{\perp}}}\cdot d\tau+\frac{1}{2}\cdot\frac{l_{a_{\eta c}}}%
{l_{\xi_{\perp}}}\cdot d\tau^{2}=d\overline{z}_{c}=\\
&  =-\frac{d\overline{z}_{c}}{d\tau}\cdot d\tau+\frac{1}{2}\cdot\frac
{d^{2}\overline{z}_{c}}{d\tau^{2}}\cdot d\tau^{2}+O(d\tau)\approx\\
&  \approx-\frac{d\overline{z}_{c}}{d\tau}\cdot d\tau+\frac{1}{2}\cdot
\frac{d^{2}\overline{z}_{c}}{d\tau^{2}}\cdot d\tau^{2}\text{\ .}%
\end{align*}

The comparison of the coefficients before $d\tau$ and $d\tau^{2}$ in the two
expressions%
\begin{align*}
z_{c}  &  =\frac{\overline{\omega}-\omega}{\omega}=-\frac{l_{v_{\eta c}}%
}{l_{\xi_{\perp}}}\cdot d\tau+\frac{1}{2}\cdot\frac{l_{a_{\eta c}}}%
{l_{\xi_{\perp}}}\cdot d\tau^{2}=d\overline{z}_{c}=\\
&  \approx-\frac{d\overline{z}_{c}}{d\tau}\cdot d\tau+\frac{1}{2}\cdot
\frac{d^{2}\overline{z}_{c}}{d\tau^{2}}\cdot d\tau^{2}\text{ \ \ }%
\end{align*}
leads to the relations%
\[
\frac{d\overline{z}_{c}}{d\tau}=\frac{l_{v_{\eta c}}}{l_{\xi_{\perp}}}\text{
\ \ \ \ , \ \ \ \ \ }\frac{d^{2}\overline{z}_{c}}{d\tau^{2}}=\frac{l_{a_{\eta
c}}}{l_{\xi_{\perp}}}\text{ \ \ .\ \ \ \ }
\]

The vector $\xi_{\perp}$ could be chosen as a unit vector, i.e. $l_{\xi
_{\perp}}=1$, equal to the vector $n_{\perp}$ showing the direction to the
emitter from point of view of the observer. Then%
\[
\frac{d\overline{z}_{c}}{d\tau}=l_{v_{\eta c}}\text{ \ \ \ \ , \ \ \ \ \ }%
\frac{d^{2}\overline{z}_{c}}{d\tau^{2}}=l_{a_{\eta c}}\text{ \ \ .\ }
\]

Therefore, if we can measure the change (variation) of the shift frequency
parameter $d\overline{z}_{c}$ in a time interval $d\tau$ we can find the
Coriolis relative velocity and Coriolis relative acceleration of the emitter
with respect to the observer. The above relations appear as direct way for a
check-up of the considered theoretical scheme of the propagation of signals in
spaces with affine connections and metrics. On the other side, the explicit
form of $l_{v_{\eta c}}$ and $l_{a_{\eta c}}$ as functions of the kinematic
characteristics of the relative velocity and relative acceleration could lead
to conclusions of the properties of the space-time model used for description
of the physical phenomena. The same relations could lead to more precise
assessment of the Hubble function $H_{c}$ and the acceleration function
$\overline{q}_{\eta c}$ at a given time.

\subsection{Hubble effect and shift frequency parameter in the case of an
auto-parallel motion of the observer}

It has been shown \cite{Manoff-8}, \cite{Manoff-9} that in a $(\overline
{L}_{n},g)$-space longitudinal (standard) and transversal Hubble effects could
appear when signals are propagating from an emitter to an observer (detector)
moving relatively to each other.

\subsubsection{Longitudinal (standard) Hubble effect and the shift frequency
parameter}

By the use of the relations \cite{Manoff-8}, \cite{Manoff-9}
\[
l_{v_{z}}=\mp H\cdot l_{\xi_{\perp}}\text{ \ \ \ \ \ \ , \ \ \ \ \ \ \ \ \ \ }%
l_{a_{z}}=\mp\overline{q}\cdot l_{\xi_{\perp}}\text{ \ \ \ \ \ ,}%
\]%
\[
\frac{d\overline{z}}{d\tau}=\frac{l_{v_{z}}}{l_{\xi_{\perp}}}\text{ \ \ \ \ ,
\ \ \ \ \ \ \ \ \ \ }\frac{d^{2}\overline{z}}{d\tau^{2}}=\frac{l_{a_{z}}%
}{l_{\xi_{\perp}}}\text{ \ \ \ \ \ \ ,}%
\]
we can find the Hubble function $H$ and the acceleration function (parameter)
$\overline{q}$ respectively as%
\[
\frac{d\overline{z}}{d\tau}=\mp H\text{ \ \ \ \ , \ \ \ \ \ \ \ }\frac
{d^{2}\overline{z}}{d\tau^{2}}=\mp\overline{q}\text{ \ \ \ \ .\ }%
\]

\subsubsection{Transversal Hubble effect and the shift frequency parameter}

By the use of the relations \cite{Manoff-8}, \cite{Manoff-9}%
\[
l_{v_{\eta c}}=\mp\overline{H}_{c}\cdot l_{\xi_{\perp}}\text{ \ \ \ \ \ ,
\ \ \ \ \ \ \ \ \ \ \ }l_{a_{\eta c}}=\mp\overline{q}_{\eta c}\cdot
l_{\xi_{\perp}}\text{ \ \ \ ,}%
\]%
\[
\frac{d\overline{z}_{c}}{d\tau}=\frac{l_{v_{\eta c}}}{l_{\xi_{\perp}}}\text{
\ \ \ \ , \ \ \ \ \ }\frac{d^{2}\overline{z}_{c}}{d\tau^{2}}=\frac{l_{a_{\eta
c}}}{l_{\xi_{\perp}}}\text{ \ \ \ ,}%
\]
we can find the transversal Hubble function $H_{c}$ and the transversal
acceleration function (parameter) $\overline{q}_{\eta c}$ respectively as%
\[
\frac{d\overline{z}_{c}}{d\tau}=\mp\overline{H}_{c}\text{ \ \ \ \ \ \ ,
\ \ \ \ \ \ \ \ \ }\frac{d^{2}\overline{z}_{c}}{d\tau^{2}}=\mp\overline
{q}_{\eta c}\text{ \ \ .}%
\]

\section{Conclusion}

In the present paper we have considered the Doppler effect and the Hubble
effect  for the case of auto-parallel motion of the observer in a
$(\overline{L}_{n},g)$-space and their relations to the shift frequency
parameters corresponding to the longitudinal and transversal effects. It is
shown that these effects lead to direct check-up of the theoretical scheme and
could be used for finding out the relative centrifugal (centripetal)
velocities and accelerations as well as the relative Coriolis velocities and
accelerations of moving astronomical objects from point of view of the proper
frame of reference of an observer (detector).

The Doppler effects and the Hubble effects are considered on the grounds of
purely kinematic considerations. It should be stressed that the Hubble
functions $H$ and $\overline{H}_{c}$ are introduced on a purely kinematic
basis related to the notions of relative centrifugal (centripetal) velocity
and to the notions of Coriolis velocities respectively. They could be found
directly by the use of the measurements of the shift frequency parameters. It
should be noted that $\overline{H}_{c}$ does not exists in the Einstein theory
of gravitation. The dynamic interpretations of $H$ and $\overline{H}_{c}$ in a
theory of gravitation depend on the structures of the theory and the relations
between the field equations and on both the functions. In this paper it is
shown that notions the specialists use to apply in theories of gravitation and
cosmological models could have a good kinematic grounds independent of any
concrete classical field theory. Doppler effects, and Hubble effects could be
used in mechanics of continuous media and in other classical field theories in
the same way as the standard Doppler effect is used in classical and special
relativistic mechanics.


\begin{thebibliography}{99}                                                                                               %
\bibitem {Weinberg}Weinberg S 1972 \textit{Gravitation and cosmology:
Principles and applications of the general theory of relativity} (New York:
John Wiley and Sons)

\bibitem {Unzicker}Unzicker A 2003 Galaxies as Rotating Buckets - a Hypothesis
on the Gravitational Constant Based on Mach's Principle. (\textit{Preprint}
ArXiv gr-qc/03 08 087)

\bibitem {Javorski}Javorskii B M and Detlaff A A 1977 \textit{Handbook of
Physics} ( Moscow: Nauka, ) (in Russian) 532 552 533

\bibitem {Tonnelat}Tonnelat M  A 1962 \textit{Foundations of electromagnetism
and the theory of relativity} \ (Moscow: Inostr. lit.) (in Russian)

\bibitem {Anderson}Anderson J A 1967\ \textit{Principles of relativity
physics} (New York: Academic Press) 270-271

\bibitem {Misner}Misner Ch W, Thorne K S and Wheeler J A 1973
\textit{Gravitation. }(San Francisco: W H Freeman and Company). Russian
translation 1977: Vol 1., Vol. 2., Vol. 3. (Moscow: Mir)

\bibitem {Manoff-1}Manoff S 2002 \textit{Geometry and Mechanics in Different
Models of Space-Time}: \textit{Geometry and Kinematics}. (New York: Nova
Science Publishers) Parts 2 - 3

\bibitem {Manoff-2}Manoff S 2002 \textit{Geometry and Mechanics in Different
Models of Space-Time}: \textit{Dynamics and Applications}. (New York:  Nova
Science Publishers) Parts 1 - 2

\bibitem {Manoff-3}Manoff S 1999 Spaces with contravariant and covariant
affine connections and metrics \textit{Physics of elementary particles and
atomic nucleus (PEPAN)} [Russian Edition: \textbf{30} 5 1211-1269], [English
Edition: \textbf{30}  5  527-549]

\bibitem {Iliev-1}Iliev B Z 1996 Normal frames and the validity of the
equivalence principle: I. Cases in a neighborhood and at a point \textit{J.
Phys. A: Math. Gen.} \textbf{29} 6895-6901

\bibitem {Iliev-1a}Iliev B Z 1997 Normal frames and the validity of the
equivalence principle: II. The case along paths \textit{J. Phys. A: Math.
Gen.} \textbf{30} 4327-4336

\bibitem {Iliev-1b}Iliev B Z 1998 Normal frames and the validity of the
equivalence principle: III. The case along smooth maps with separable points
of self-interaction \textit{J. Phys. A: Math. Gen.} \textbf{31} 1287-1296

\bibitem {Manoff-5}Manoff S 2000 Fermi derivative and Fermi-Walker transports
over $(\overline{L}_{n},g)$-spaces \textit{Intern. J. Mod. Phys.} \textbf{A
13}  5  679-695

\bibitem {Manoff-6}Manoff S 1998 Conformal derivative and conformal transports
over $(\overline{L}_{n},g)$-spaces \textit{Intern. J. Mod. Phys}. \textbf{A
15}  25 4289-4308

\bibitem {Manoff-7}Manoff S 2002 Mechanics of continuous media in
$(\overline{L}_{n},g)$-spaces. 1. Introduction and mathematical tools.
(\textit{Preprint ArXiv}: gr-qc / 02 03 016

\bibitem {Manoff-7a}Manoff S 2002 Mechanics of continuous media in
$(\overline{L}_{n},g)$-spaces. 2. Relative velocity and deformations.
(\textit{Preprint ArXiv}: gr-qc / 02 04 003

\bibitem {Manoff-7b}Manoff S 2002 Mechanics of continuous media in
$(\overline{L}_{n},g)$-spaces. 3. Relative accelerations. (\textit{Preprint
ArXiv}: gr-qc / 02 03 016

\bibitem {Manoff-8a}Manoff S 2003 Centrifugal (centripetal), Coriolis
velocities, accelerations, and Hubble law in spaces with affine connections
and metrics (\textit{Preprint ArXiv}: gr-qc / 02 12 038); \textit{Central
European J. of Physics} \textbf{4} 669-694

\bibitem {Manoff-8}Manoff S 2003 Propagation of signals in spaces with affine
connections and metrics (\textit{Preprint ArXiv}: gr-qc / 03 09 050)

\bibitem {Manoff-9}Manoff S 2004 \textit{e}Propagation of signals in spaces
with affine connections and metrics as models of space-time [to appear in
\textit{Physics of elementary particles and atomic nuclei (PEPAN}), \textbf{4} (2004)]

\bibitem {Stephani}Stephani H 1977 \textit{Allgemeine Relativitaetstheorie}
(Berlin: VEB Deutscher Verlag d. Wissenschaften) 75-76

\bibitem {Manoff-12}Manoff S 2001 Frames of reference in spaces with affine
connections and metrics (\textit{Preprint ArXiv} gr-qc/99 08 061)
\textit{Class. Quantum Grav.} \textbf{18} 6 1111-1125 

\bibitem {Manoff-11}Manoff S 1995 Kinematics of vector fields In
\textit{Complex Structures and Vector Fields}. eds. Dimiev St., Sekigawa K.
(Singapore: World Scientific )  61-113
\end{thebibliography}
\end{document}